\documentclass[prl,twocolumn,showpacs,preprintnumbers,amsmath,amssymb]{revtex4}

\newcommand{\Aphys}{-17.23}              
\newcommand{\Deltastat}{0.82}            
\newcommand{\Deltasyst}{0.89}            
\newcommand{\Azero}{-15.87}              
\newcommand{\DeltaAzero}{1.22}           
\newcommand{\FakGEs}{0.26}               
\newcommand{\GEsGMs}{-0.12 }              
\newcommand{\DeltaGEsGMs}{$0.11\pm0.11$}         
\newcommand{\BeamEnerg} {315.1}          
\newcommand{\Qsquareshort}               {0.22} 
\newcommand{\Qsquare} {\Qsquareshort~(GeV/$c$)$^2$} 
\newcommand{\GEs}{0.050}                 
\newcommand{\DeltaGEsexp}{0.038}        
\newcommand{\DeltaGEsff}{0.019}         
\newcommand{\GMs}{-0.14}            
\newcommand{\DeltaGMsexp}{0.11}        
\newcommand{\DeltaGMsff}{0.11}         
\newcommand{\GEscomb}{0.035}                 
\newcommand{\DeltaGEscombexp}{0.030}        
\newcommand{\DeltaGEscombff}{0.019}         

\newcommand{\ppm}{$\cdot 10^{-6}$}       
\usepackage{graphicx}
\usepackage{dcolumn}
\usepackage{bm}

\begin{document}

\preprint{APS/123-QED}

\title{Measurement of Strange Quark Contributions to the Vector Form Factors of the Proton at $Q^2$=\Qsquare}
\author{S.~Baunack$^*$}
\author{K.~Aulenbacher}
\author{D.~Balaguer R\'{i}os}
\author{L.~Capozza} 
\author{J.~Diefenbach}
\author{B.~Gl\"aser}
\author{D.~von~Harrach}
\author{Y.~Imai}
\author{E.-M.~Kabu\ss}
\author{R.~Kothe}
\author{J.~H.~Lee}
\author{H.~Merkel}
\author{M.~C.~Mora~Esp\'{i}}
\author{U.~M{\"u}ller}
\author{E.~Schilling}
\author{G.~Stephan}
\author{C.~Weinrich}
\affiliation{Institut f\"ur Kernphysik, Johannes Gutenberg-Universit\"at Mainz,
  J.J.~Becherweg~45, D-55099~Mainz, Germany
}

\author{J.~Arvieux$^\dagger$}
\author{M.~A.~El-Yakoubi}
\author{R.~Frascaria}
\author{R.~Kunne}
\author{F.~E.~Maas$^\ddagger$}
\author{M.~Morlet}
\author{S.~Ong}
\author{J.~van~de~Wiele}
\affiliation{
Institut de Physique Nucl\'{e}aire, CNRS-IN2P3, Universit\'{e} Paris-Sud, F-91406 Orsay Cedex, France
}

\author{S.~Kowalski}
\author{Y.~Prok}
\author{S.~Taylor}
\affiliation{
Laboratory for Nuclear Science and Department of Physics, Massachusetts Institute of Technology,
Cambridge, MA~02139, USA
}

\date{\today}

\begin{abstract}
A new measurement of the parity violating asymmetry in elastic
electron scattering on hydrogen at backward angles and at a four momentum
transfer of $Q^2=$\Qsquare~is reported here. The measured asymmetry is $A_{\rm
  LR}=(\Aphys \pm
\Deltastat_{\rm stat} \pm \Deltasyst_{\rm syst})$\ppm. The Standard Model prediction
assuming no strangeness is $A_0=(\Azero \pm \DeltaAzero)$~\ppm. In combination with previous
results from measurements at forward angles, it it possible to disentangle for the first time the strange electric and magnetic form
factors at this momentum transfer, $G_E^s($\Qsquare$)=\GEs\pm\DeltaGEsexp\pm\DeltaGEsff$
and $G_M^s($\Qsquare$)=\GMs\pm\DeltaGMsexp\pm\DeltaGMsff$.
\end{abstract}

\maketitle
Sea quarks are an important ingredient to describe nucleon properties in terms of fundamental
QCD degrees of freedom. Strange quark-antiquark pairs might play a
relevant role and affect e.g. the electromagnetic properties of the
nucleon. The contribution of strange quarks to the charge radius
and magnetic moment in the nucleon ground state is of specific
interest since this is a pure sea quark effect.
The strange quark contribution to the electromagnetic form factors of the nucleon can be expressed
in terms of the strange electric and magnetic form factors $G_E^s$ and
$G_M^s$. There are various theoretical approaches for estimating the strange form factors~\cite{PAVI04,PAVI06},
such as quark soliton models \cite{Silva06,Goeke07,Weigel95}, chiral quark
models \cite{Lyubovitskij02}, quenched lattice
calculations~\cite{Leinweber08} or two-component models~\cite{Bijker06}. Parity violating electron scattering provides
a direct experimental approach~\cite{KaplanManohar88, Musolf94, BeckHols}.\\
A measurement of parity violation necessarily involves a weak interaction
probe of the nucleon. This provides additional information allowing a
measurement of $G_E^s$ and $G_M^s$. Within
the standard model of electroweak interaction, it is known that electromagnetic and weak currents are
related. Assuming isospin symmetry, the weak vector form factors $\tilde{G}^p_{E,M}$ of the
proton, describing the vector coupling to the $Z^0$ boson, can be
expressed in terms of the electromagnetic nucleon form factors
$G_{E,M}^{p,n}$ and the strange form factors $G^s_{E,M}$. The
interference between tree level electromagnetic and weak amplitudes
leads to a parity violating asymmetry in the elastic scattering cross
section of left- and right-handed electrons (LR) $\sigma^L$, $\sigma^R$:
$A_{\rm LR}=(\sigma^R-\sigma^L)/(\sigma^R+\sigma^L)$.
This asymmetry can be written as a sum of three terms,
$A_{\rm LR}=A_{\rm V}+A_{\rm S}+A_{\rm A}$. $A_{\rm V}$ represents the vector coupling on
the proton vertex without strangeness contribution, $A_{\rm S}$ contains the strange quark
vector contribution, and $A_{\rm A}$ represents the axial coupling to the proton
vertex \cite{Musolf94}:
\begin{eqnarray}
A_{\rm V}=-a \rho'_{eq}\left[(1-4\hat{\kappa}'_{eq}\hat{s}^2_Z)
     -\frac{\epsilon G_E^p G_E^n + \tau G_M^p G_M^n}{\epsilon (G_E^p)^2 + \tau
     (G_M^p)^2} \right] \label{eq:Av} \\
A_{\rm S}=a{\rho'_{eq}\frac{\epsilon G_E^p G_E^s + \tau G_M^p G_M^s}{\epsilon
(G_E^p)^2 + \tau (G_M^p)^2}} \label{eq:As}\\
A_{\rm A}=a{\frac{(1-4\hat{s}^2_Z)\sqrt{1-\epsilon^2}{\sqrt{\tau(1+\tau)}G_M^p\tilde{G}_A^p}}{\epsilon
(G_E^p)^2 + \tau (G_M^p)^2}} \label{eq:Aa} 
\end{eqnarray} 
with $a=\frac{G_{\mu}Q^2}{4 \pi \alpha \sqrt{2}}$, $G_{\mu}$ the Fermi
coupling constant, $\alpha$ the fine structure constant,
$\tau=Q^2/(4 M_{\rm p}^2)$, $Q^2$ the negative squared four momentum
transfer, $M_{\rm p}$ the proton mass and $\epsilon=[1+2(1+\tau)\tan^2
\frac{\Theta}{2}]^{-1}$. $\Theta$ is the scattering angle in the
laboratory frame and $\hat{s}_Z^2(M_Z)=0.23119(14)$~\cite{PDG08} is the
square of the sine of the weak-mixing angle. The factors $\rho_{eq}'$ and
$\hat{\kappa}'_{eq}$ include the electroweak radiative corrections
evaluated in the minimal subtraction renormalization scheme ($\overline{MS}$). The electromagnetic form factors
$G_{E,M}^{p,n}$ are taken from a Monte Carlo based analysis of the world
data~\cite{Elyakoubi07} resulting in the non-strangeness expectation $A_0=A_V+A_A=
(\Azero \pm \DeltaAzero)$~\ppm. The dominant contribution to the uncertainty
of $A_0$ comes from the uncertainty due to the two-quark radiative corrections
(anapole moment) in the axial form factor $\tilde{G}^p_A$, followed by the uncertainties in $G_M^n$ and $G_M^p$.\\
Recently four groups have published related
results. The SAMPLE collaboration at MIT-Bates
\cite{Spayde04, Ito04} involved a backward angle measurement on a
hydrogen target at a four momentum transfer of $Q^2=0.1~$(GeV$/c)^2$
and on deuterium at $Q^2=0.1~($GeV$/c)^2$ and
$0.04~($GeV$/c)^2$, being sensitive mainly to $G_M^s$ and $\tilde{G}^p_A$.
The HAPPEX collaboration at TJNAF reported a
measurement on a hydrogen target at forward angles at a $Q^2$ of
$0.47~($GeV$/c)^2$, mainly sensitive to $G_E^s$ \cite{Happex03} and a precise
measurement with helium and proton targets at a $Q^2$ of $0.1~($GeV$/c)^2$
\cite{Happex07}. Those measurements put tight constraints on the strangeness contribution to the form factors at
these momentum transfers. 
The G0 collaboration at TJNAF performed a forward angle measurement with
several momentum transfers between $0.1~($GeV$/c)^2$ and
$1~($GeV$/c)^2$, including the momentum transfer discussed here~\cite{Armstrong05}.
The A4 collaboration at MAMI has completed measurements on
a hydrogen target at forward angles and momentum transfers of $0.23~($GeV$/c)^2$
and $0.1~($GeV$/c)^2$  \cite{Maas04, Maas05}. These measurements were sensitive mainly to $G_E^s$.
Here, a new measurement of the parity violating asymmetry in
the elastic scattering of polarized electrons off unpolarized protons
at backward angles at 
$Q^2=$~\Qsquare~is presented and the implications of the new result for the
strange form factors are discussed.
A single measurement of $A_{\rm S}$ gives a linear combination of $G_E^s$ and
$G_M^s$. Combining two measurements with different kinematics allows
the separation of the two strange form factors when $\tilde{G}_A^p$ is taken from a calculation as an input parameter.
These two measurements have to be performed at the same $Q^2$. 
The A4 experimental setup~\cite{Maas03,Maas04} at the MAMI
accelerator~\cite{MAMI} allows the measurement at scattering angles $\Theta$
either between 30$^\circ$ and 40$^\circ$ or 140$^\circ$ and 150$^\circ$. As
$Q^2=4 \eta^{-1}E^2 \sin^2\left(\Theta/2\right)$, where $\eta = 1 + E/M_{\rm p}(1-\cos\Theta)$,
the momentum transfer can be selected by varying the beam energy $E$. To
match the $Q^2$ value of the A4 forward
measurement a beam energy of \BeamEnerg~MeV was chosen for the backward angle experiment.\\
%
The experimental setup was described in detail in \cite{Maas04}. Here we
summarize the basic components and emphasize the modifications that have been made.
A superlattice photocathode delivered a polarized electron beam with an intensity of
$20~\mu$A and an average polarization $P_e$ of about 70\%. The beam polarization was measured once a week
using a M{\o}ller polarimeter with a precision of 2\%. In addition, 
a Mott polarimeter and a transmission Compton polarimeter were
used. Altogether, the uncertainty to the beam polarization is 4\%.
%
It was particularly important to minimize helicity correlated beam fluctuations in
position, angle, current and energy that introduce false asymmetries due to
changes in luminosity, cross section or solid angle. Table
\ref{tab:Paras_315MeV} lists the measured beam parameters during the
1100 hours of asymmetry data taking.
\begin{table}
   \caption{\label{tab:Paras_315MeV} Average beam parameters $\bar{X}_i$ of the electron beam and the associated asymmetries $a_i\bar{X}_i$ as defined in eq.~\ref{gl:Ansatz}.}
   \begin{ruledtabular}
    \begin{tabular}{l l | r l | r l} 
      $i$  & Parameter                 & $\bar{X}_i$&&$a_i\cdot \bar{X}i$&\\ \hline
      1  & Current Asymmetry         & $-0.30$&ppm&-0.25&ppm \\ 
      2  & Horizontal Position Diff.  & $-86.97$&nm&+0.61&ppm\\ 
      3  & Vertical Position Diff.  & $-23.84$&nm&-0.86&ppm  \\ 
      4  & Horizontal Angle Diff.     & $-8.53$&nrad&-0.09&ppm \\ 
      5  & Vertical Angle Diff.     & $-2.40$&nrad&+0.10&ppm\\ 
      6  & Energy Diff.         & $-0.41$&eV&+0.16&ppm\\ 
    \end{tabular}   
  \end{ruledtabular}  
\end{table}
The liquid hydrogen target~\cite{Altarev05} was 23.4~cm long yielding a luminosity
$L\approx1.2\cdot10^{38}$~cm$^{-2}$s$^{-1}$.  Target density
fluctuations were monitored by eight water-Cherenkov luminosity
monitors located at small scattering angles~\cite{Hammel05} and were kept
below $\Delta L/L<10^{-6}$ averaged over the whole data set.
The scattered electrons were detected in a homogenous
electromagnetic calorimeter that consists of 1022 lead
fluoride (PbF$_2$) crystals~\cite{Achenbach01}. The detector covered a solid angle of $\Delta\Omega$=$0.62$~sr.
Single events were detected and their energy was measured with a resolution of about $3.9\%/\sqrt{E}$.\\
\begin{figure}
\includegraphics[width=0.4\textwidth]{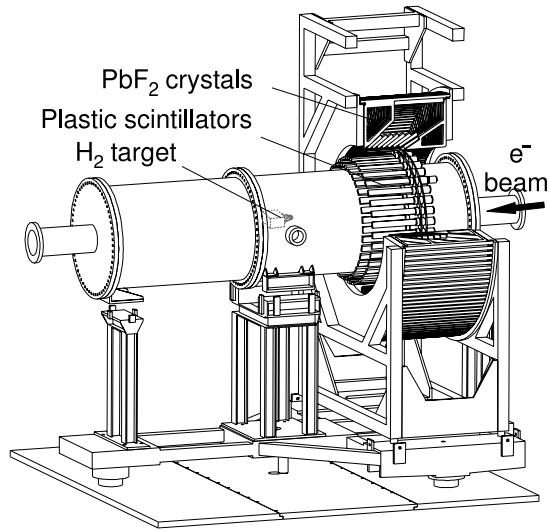}
\caption{\label{fig:DetFig}Drawing of the PbF$_2$ calorimeter together with
  the scattering chamber. The scintillators are
  placed between the scattering chamber and the lead fluoride crystals. The
  whole system is mounted on a rotatable platform so that both forward and
  backward angle configuration can be easily set up.}
\end{figure}
The most important aspect of the backward measurement is the
installation of 72 plastic scintillators in front of the
PbF$_2$ crystals (see Fig. \ref{fig:DetFig}). Used in coincidence with
the calorimeter, they enable the separation of charged from
neutral particles. Photons from $\pi^0$ decay could
thus be separated from scattered electrons.
If the electronic threshold was exceeded, the energy that was deposited by a particle
in the calorimeter was digitized by an 8 bit ADC and stored into a coincidence
or a noncoincidence histogram depending on the trigger signal from the
scintillator. Furthermore a polarization signal distinguished between the two
helicity states of the beam.    
Altogether each calorimeter channel produced four histograms for each five minute data taking run.\\
The data analysis was similar to that of the previous
measurements~\cite{Maas05}. Modifications were needed since the coincidence
histograms were polluted by high energy photons converting into e$^+$e$^-$ pairs
in the aluminium wall of the vacuum chamber and in the scintillator. The LR-asymmetry of the
$\gamma$ background was determined from the noncoincidence spectra. A detailed
Monte Carlo simulation using {\small GEANT4} was implemented
for tracking shower particles and calculating the detector
response. The simulation reproduced the measured spectrum well for
energies above 125~MeV, while for lower energies threshold effects of the
analogue readout electronics become important. From the simulation one can
derive as a function of the $\gamma$ energy both the probability of a $\gamma$ to convert and trigger the
scintillator and the mean energy loss of the generated e$^+$e$^-$ pairs.
\begin{figure}
  \begin{center}
    \includegraphics[width=0.50\textwidth]{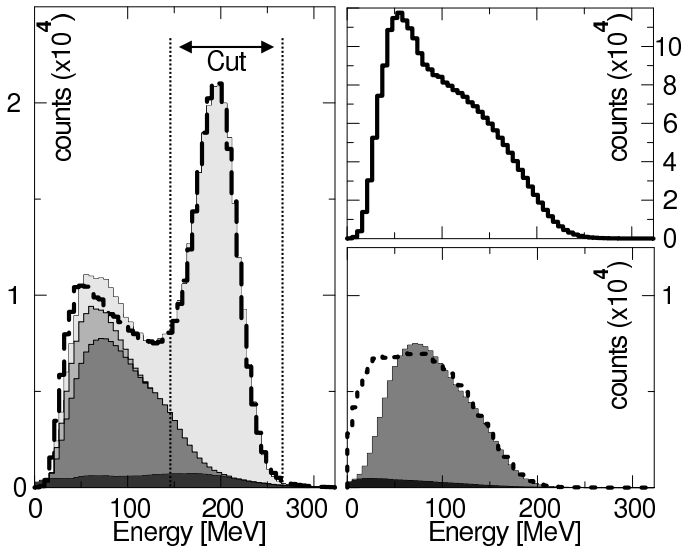}
    \caption{\label{fig:Simulation} Left panel: The measured energy spectrum of
      coincidence events is shown by the dashed line. One can clearly identify the peak of
      the elastic scattered electrons. The contributions of the different
      processes are shown from bright to dark: (i) the elastically scattered
      electrons, (ii) the inelastically scattered electrons, (iii)
      the converted photons from $\pi^0$-decay, and (iv) empty target background.
      Upper right panel: The solid line shows a measured energy
      spectrum of noncoincidence events. Lower right panel: The
      dotted line shows the background contribution to the coincidence spectrum estimated from the
      noncoincidence events by applying the shifting and scaling method in
      comparison with the photon background obtained from the simulation
      (gray) together with the shifted and scaled noncoincidence events from
      an empty target measurement (dark).}
  \end{center}
\end{figure}
Fig.~\ref{fig:Simulation} shows the measured energy spectra and the contributions from the different
processes. The contribution to the background arising from aluminium events from the
target entrance- and exit-windows was determined by a measurement with an
empty target and is about 4.5\%.
The background elimination was achieved by scaling the measured noncoincidence spectra
with the conversion probability and shifting them by the energy loss. Different methods
for the scaling and shifting procedure were applied and gave differences in the
final asymmetry below $0.2\cdot10^{-6}$ .\\
\begin{figure}
\includegraphics[width=0.5\textwidth]{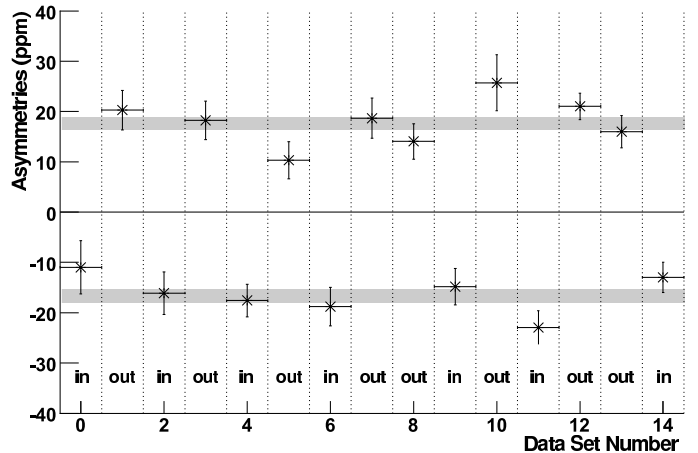}
\caption{\label{fig:SamplePlot}Measured asymmetries $A_{\rm LR}$ with
  respect to the position of the half-wave plate at the electron source
  (IN/OUT). The reversal of the helicity can be easily observed in the sign flip
  of the extracted asymmetries. The two grey bands show fits to the data,
  $A_{\rm OUT}=(17.70\pm1.27)$~ppm and $A_{\rm IN}=(-16.68\pm1.37)$~ppm (not all
  systematic errors included here).} 
 \end{figure}
\begin{table}
  \caption{\label{tab:Korrekturen570MeV}Applied corrections to the measured asymmetry and their contribution to the systematic error.}
  
    \begin{tabular}{|l|r r|}\hline
       Applied scaling factor           & Scaling factor  & Error\\\hline
       Polarization $P_e$               &  0.683          & 0.040 \\\hline\hline
       Applied corrections              & Correction      & Error \\ 
                                       & (ppm)           & (ppm) \\\hline
      Helicity correlated beam differences  &  0.14           & 0.39 \\
      Accidental coincidence events    & -0.19           & 0.02 \\
      Al windows (H$_2$ target)        &  0.29           & 0.04 \\
      Dilution of $\pi^0$ decay photons& -1.49           & 0.28 \\\hline
    \end{tabular}
\end{table}
The number of elastic events for positive and negative helicity
was determined by applying cuts on the coincidence
energy histograms as indicated in Fig.~\ref{fig:Simulation} by the dotted lines
and summing up all 730 channels of the inner
five calorimeter rings. For each run the raw asymmetry is calculated. The false
asymmetries are corrected using the ansatz\vspace{-0.3cm}
\begin{equation}
\label{gl:Ansatz}
A_{\rm raw}=A_{\rm exp}+\sum_{i=1}^{6}{a_i X_i}
\vspace{-0.2cm}
\end{equation}
where the $X_i$ denote the helicity correlated beam parameters as
defined in table \ref{tab:Paras_315MeV}. The $a_i$
denote the correlation coefficients between the observed asymmetry $A_{\rm raw}$
and the beam parameters $X_i$. These coefficients have been
determined from geometry and in addition from the intrinsic beam
fluctuations via a multiple linear regression analysis. Both
methods yield only small corrections relative to the measured asymmetry and
agree within the statistical precision. Finally the physical asymmetry $A_{\rm
  LR}$ is obtained
by normalization of $A_{\rm exp}$ by the electron beam polarization $P_{\rm
  e}$: $A_{\rm LR}=A_{\rm exp}/P_{\rm e}$.
About half of our data was taken with a half-wave plate inserted in the
laser optics of the electron source. This leads to a reversal of the beam
helicity and a partial compensation of helicity correlated false
asymmetries. All relevant corrections applied to the measured asymmetry are
listed in table \ref{tab:Korrekturen570MeV}. The asymmetry for the aluminium events is
calculated in the static approximation leading to a correction for the asymmetry of 0.29\ppm. Another
source of background are accidental coincidence events in the scintillators with a fraction
of about 1.3\% leading to a correction of $-0.19$\ppm. Since the event rate on
the detector is 4-8 times smaller than in our forward measurements,
corrections on the asymmetry due to pile-up are negligible here.
\begin{figure}[t]
\includegraphics[width=0.5\textwidth]{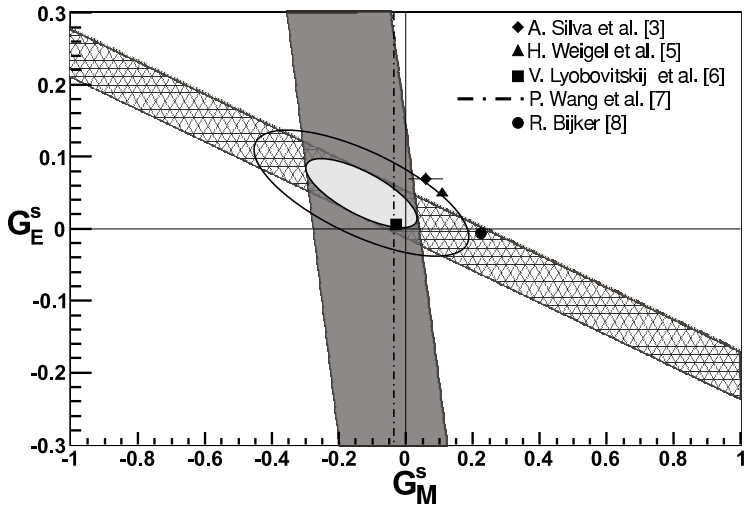}
\caption{\label{fig:GeGm} The linear combination of $G_E^s+\zeta G_M^s$
  as extracted from our measurement presented here (the solid band) together with
  the A4  forward angle measurement \cite{Maas04} (hatched band) at the same
  momentum transfer. The bands represent the
  possible values of $G_E^s~+~\zeta G_M^s$ within the one-$\sigma$-uncertainty
  with statistical and systematic error added in quadrature. The ellipses show
  the 68\% and 95\% C.L. constraints in the $G_E^s$-$G_M^s$ plane. Also shown
  are theoretical predictions \cite{Silva06,Weigel95,Lyubovitskij02,Leinweber08, Bijker06}.
}
\end{figure}
Fig. \ref{fig:SamplePlot} shows the parity violating asymmetries
for the whole data set. The sign flip when the half-wave plate was inserted can be
clearly observed. In total $3\cdot10^{12}$ coincidence events were used for the full analysis.
An asymmetry of $A_{\rm LR}=(\Aphys\pm\Deltastat_{\rm stat}\pm\Deltasyst_{\rm syst})$\ppm~is extracted.\\
From the difference between $A_{\rm LR}$ and the asymmetry without strangeness $A_0$ 
the linear combination of the strange electric and magnetic form factors
$G_M^s+\FakGEs~G_E^s=\GEsGMs\pm$\DeltaGEsGMs~is obtained,
where the first error comes from the measurement and the second from the
uncertainty in the axial and electromagnetic form factors of the nucleon. In
Fig.~\ref{fig:GeGm} the shaded band shows the possible values of $G_E^s$ and
$G_M^s$ within the one-$\sigma$-uncertainty. The hatched band shows 
the A4 result of the forward angle measurement at the same momentum transfer. 
Due to a careful re-analysis of the electron polarization measurement and
using an up-to-date parameterization of the electromagnetic form factors~\cite{Elyakoubi07}, the value of the linear
combination has shifted down from $G_E^s+0.225~G_M^s=0.039\pm0.028\pm0.020$
as presented in \cite{Maas04} to $G_E^s+0.224~G_M^s=0.020\pm0.029\pm0.016$. Disentangling the linear combinations, one
gets $G_E^s=\GEs\pm\DeltaGEsexp\pm\DeltaGEsff$ and $G_M^s=\GMs\pm\DeltaGMsexp\pm\DeltaGMsff$. 
A combined analysis including the G0 forward angle measurement at $Q^2=0.23$~GeV$^2$ results in a more precise value for $G_E^s$, namely
$G_E^s=\GEscomb\pm\DeltaGEscombexp\pm\DeltaGEscombff$.\\
The strange form factors presented here are determined simultaneously from two
complementary A4 measurements with the same momentum transfer using the same
method. In contrast to the only existing published backward angle measurement at a lower $Q^2$ of
0.1~GeV/$c^2$ favoring a positive value of $G_M^s$~\cite{Spayde04}, the new
result favors a negative strange magnetic moment as predicted by many models
and also in accordance with the latest lattice calculation~\cite{Leinweber08}.
Furthermore, it disfavors a negative $G_E^s$ in this momentum transfer region as suggested by~\cite{Armstrong05}. Both HAPPEX and A4
have scheduled measurements in the near future to clarify the
situation for $G_E^s$ at $Q^2=0.6$~GeV/$c^2$.
\bibliographystyle{prsty}
\bibliography{315H2back}
\end{document}